\documentclass[11pt, draft]{amsart}

\usepackage{amssymb}

\newcommand{\ci}[1]{_{ {}_{\scriptstyle #1}}}
\newtheorem*{theorem}{Theorem}
\newtheorem*{corollary}{Corollary}
\newtheorem*{Koosisthm}{Theorem of Koosis}

\renewcommand{\le}{\leqslant}
\renewcommand{\ge}{\geqslant}

\newcommand{\C}{{\mathbb C}}
\newcommand{\T}{{\mathbb T}}
\newcommand{\D}{{\mathbb D}}
\newcommand{\cP}{{\mathcal P}}

\newcommand{\ds}{\displaystyle}

\title[ Asymptotics of orthogonal polynomials via the Koosis theorem ]{
Asymptotics of orthogonal polynomials via the Koosis theorem  }
\author{ F. Nazarov, A. Volberg, P. Yuditskii }

\thanks{Partially supported by NSF grant DMS-0200713}

\begin{document}

\begin{abstract}
The main aim of this short paper is to advertize the Koosis theorem in
the
mathematical community, especially among those who study orthogonal
polynomials.
We (try to) do this by proving a new theorem about asymptotics of
orthogonal
polynomials for which the Koosis theorem seems to be the most natural
tool. Namely,
we consider the case when a Szeg\"o measure on the unit circumference is
perturbed
by an arbitrary measure inside the unit disk and an arbitrary Blaschke
sequence
of point masses outside the unit disk.
\end{abstract}

\maketitle

\section{Introduction and statement of results}
\label{Intr}

Consider a measure $\mu$ on the complex plane $\C$ of the form
$\mu=\nu+w\,dm+\sum_k \mu_k\delta_{z_k}$ where $\nu$ is an
arbitrary finite measure in the open unit disk $\D=\{z\in\C\,:\,|z|<1\}$,
$m$ is the Haar
measure on the unit circumference $\T$, $w\in L^1(m)$ is a strictly
positive function
satisfying the Szeg\"o condition $\int_{\T}\log w\,dm>-\infty$, $\mu_k>0$
satisfy
$\sum_k\mu_k<+\infty$, and, at last, the points $z_k$ are taken in the
exterior of the
unit disk, i.e., $|z_k|>1$ for each $k$, and satisfy the Blaschke
condition
$\sum_k(|z_k|-1)<+\infty$.

Let
$$
p_n(z)=\tau_n z^n+\dots
$$
be the $n$-th orthogonal polynomial with
respect to the
measure $\mu$ normalized by the conditions $\|p_n\|\ci{L^2(\mu)}=1$,
$\tau_n>0$.

\begin{theorem}
\label{tau}
$$\lim_{n\to\infty}\tau_n=\exp\left\{-\frac12\int_{\T}\log
w\,dm\right\}\prod_k\frac{1}{|z_k|}\,.$$
\end{theorem}

Let us introduce two auxiliary functions: the outer function
$\psi$ in the exterior
of the unit disk such that $|\psi|^2=\dfrac 1w$ on $\T$ and
$\psi(\infty)>0$, and the Blaschke product
$\ds B(z)=\prod_k \frac{\bar z_k}{|z_k|}\frac{z-z_k}{z\bar z_k-1}$.

\begin{corollary}
\label{asymp}
For every $z\in\C$ with $|z|>1$, we have
$$
\lim_{n\to \infty} \frac{p_n(z)}{z^n} = (B\psi)(z)\,.
$$
\end{corollary}

A few words about the history of the problem.
For finitely many point masses lying on the real line the theorem was
proved by Nikishin \cite{Niki}. Nikishin's result
has been generalized in various ways by Benzin and Kaliagin {\cite{BK} and
by Li and Pan in \cite{LP}.
Peherstorfer and Yuditskii \cite{PY} seem to be the first to consider the
case of infinitely many point masses.
They proved the theorem for the case when all masses lie on the real line.
An interesting attempt to deal with the
general case was made by Peherstorfer, Volberg, and Yuditskii in
\cite{PVY}.  There the analog of Theorem \ref{tau} was proved for the asymptotic of orthogonal system of {\it rational functions} (in the spirit of CMV matrices). Whether this approach can give the asymptotic of orthogonal {\it polynomials} is not clear to us at this moment.

\section{Proof of Theorem}
\label{taun}

Let us show first that $\limsup_{n\to\infty}\tau_n$ does not exceed the
right hand side.
To this end, let us observe that the right
hand side can be rewritten as $\psi(\infty)B(\infty)$ where, as before
$\psi$ is the outer function in the exterior
of the unit disk such that $|\psi|^2=\dfrac 1w$ on $\T$ and
$\psi(\infty)>0$, and
$\ds B(z)=\prod_k \frac{\bar z_k}{|z_k|}\frac{z-z_k}{z\bar z_k-1}$
is the Blashke product with zeroes $z_k$.
Now fix $\ell>0$
and put
$\ds B_\ell(z)=\prod_{k\,:\,k\le\ell} \frac{\bar
z_k}{|z_k|}\frac{z-z_k}{z\bar z_k-1}$.
Consider the integral $\ds\int_{\T}\frac{p_n}{z^n\psi B_\ell}\,dm$. On
one hand, its absolute value
does not exceed
$$
\int_{\T}\frac{|p_n|}{|\psi|}\,dm\le
\Bigl(\int_{\T}\frac{|p_n|^2}{|\psi|^2}\,dm\Bigr)^{\frac12}
=\|p_n\|\ci{L^2(w\,dm)}\le \|p_n\|\ci{L^2(\mu)}=1\,.
$$
On the other hand, this integral can be easily computed using the residue
theorem. It equals
$$
\frac{\tau_n}{\psi(\infty)B_\ell(\infty)}-\sum_{k\,:\,k\le
\ell}\frac{p_n(z_k)}{z_k^{n+1}\psi(z_k)B'_\ell(z_k)}\,.
$$
Note now that $|p_n(z_k)|\le
\mu_k^{-\frac12}\|p_n\|\ci{L^2(\mu)}=\mu_k^{-\frac12}$ for any $n$ and
$z_k^{n+1}\to\infty$
as $n\to\infty$. Therefore, $\ds \sum_{k\,:\,k\le
\ell}\frac{p_n(z_k)}{z_k^{n+1}\psi(z_k)B'_\ell(z_k)}\to 0$ as $n\to\infty$
and we conclude that $\limsup_{n\to\infty}\tau_n\le
\psi(\infty)B_\ell(\infty)$. Since it is true for any $\ell$, we can
pass to the limit as $\ell\to\infty$ and get
$\limsup_{n\to\infty}\tau_n\le \psi(\infty)B(\infty)$.

Now let us prove that $\liminf_{n\to\infty}\tau_n\ge
\psi(\infty)B(\infty)$. To this end, observe that
$$
\tau_n=\sup\{\tau\,:\,\text{there exists }p(z)=\tau z^n+\dots\text{ with
}\|p\|\ci{L^2(\mu)}\le 1\}\,.
$$
This means that is would suffice to construct a sequence of polynomials
$q_n$ such that the leading coefficients of
$q_n$ are arbitrarily close to $\psi(\infty)B(\infty)$ and
$\limsup_{n\to\infty}\|q_n\|\ci{L^2(\mu)}\le 1$.

The construction is extremely easy and well known when $\psi, B\in
C^\infty(\T)$, which corresponds to the case
when $w\in C^\infty(\T)$ and $B$ is a finite Blaschke product. In this
case all one needs to do is to expand the
analytic (in the exterior of the unit disk) function $F(z)=\psi(z)B(z)$
into its Taylor series at infinity:
$F(z)=\tau_0+\tau_1 z^{-1}+\tau_2 z^{-2}+\dots$ and put $q_n(z)=z^n
S_n(z)$ where $S_n(z)=\sum_{j=0}^n \tau_j z^{-j}$
is the $n$-th partial sum of this series. Clearly, the leading coefficient
of $q_n$ is exactly
$\tau_0=F(\infty)=\psi(\infty)B(\infty)$ for all $n$. On the other hand,
since $F\in C^{\infty}(\T)$, the partial sums
$S_n$ converge to $F$ uniformly on $\T$, which allows to estimate the
norms $\|q_n\|\ci{L^2(\mu)}$ as follows.

First of all,
$$
\int_{\T} |q_n|^2w\,dm=\int_{\T} |S_n|^2w\,dm\to \int_{\T} |F|^2w\,dm=1\,.
$$
Secondly, for each $k$,
$$
|q_n(z_k)|=|z_k^nS_n(z_k)|=|z_k^n(S_n(z_k)-F(z_k))|\le \max_\T |S_n-F|\to
0\text{ as }n\to\infty
$$
(the last inequality is just the maximum principle for the bounded and
analytic in the exterior of the unit disk
function $z^n(S_n(z)-F(z))=-\tau_{n+1}z^{-1}-\tau_{n+2}z^{-2}-\dots$).
Thus, $\sum_k\mu_k|q_n(z_k)|^2\to 0$ as $n\to\infty$
(let us remind the reader that the sum is assumed to be finite here).

Thirdly and finally, for every $z\in\D$, we have $|q_n(z)|\le
\max_\T|q_n|=\max_\T |S_n|\to\max_\T|F|$ as $n\to\infty$, so the functions
$q_n$ are uniformly bounded in $\D$. On the other hand, it is fairly easy
to see that, for any $\ell^2$ sequence
$\{\tau_j\}_{j\ge 0}$ and any $z\in\D$, the sequence $\sum_{j=0}^n\tau_j
z^{n-j}$ ($n=1,2,\dots$) tends to $0$ as
$n\to\infty$ and, moreover, this convergence is uniform on any compact
subset of $\D$. Indeed, we have
\begin{multline*}
\Bigl|\sum_{j=0}^n\tau_j z^{n-j}\Bigr|\le \sum_{0\le j\le \frac
n2}|\tau_j|\cdot|z|^{n-j}+
\sum_{\frac n2< j\le n}|\tau_j|\cdot|z|^{n-j}
\\
\le \Bigl(\sum_{j\ge 0}|\tau_j|^2\Bigr)^{\frac12}\cdot\Bigl(\sum_{0\le
j\le \frac n2}|z|^{2n-2j}\Bigr)^{\frac 12}
+\Bigl(\sum_{j> \frac n2}|\tau_j|^2\Bigr)^{\frac12}\cdot\Bigl(\sum_{0\le
j\le n}|z|^{2n-2j}\Bigr)^{\frac 12}
\\
\le \Bigl(\sum_{j\ge
0}|\tau_j|^2\Bigr)^{\frac12}\cdot\left(\frac{|z|^n}{1-|z|^2}\right)^{\frac
12}+
\Bigl(\sum_{j> \frac
n2}|\tau_j|^2\Bigr)^{\frac12}\cdot\left(\frac{1}{1-|z|^2}\right)^{\frac
12}
\end{multline*}
It remains to note that $|z|^n\to 0$ and $\sum_{j>\frac n2}|\tau_j|^2\to
0$ as $n\to\infty$. Thus, $q_n(z)\to 0$
uniformly on compact subsets of $\D$ as $n\to\infty$ and, by the dominated
convergence theorem, $\int_\D |q_n|^2\,d\nu\to 0$
as $n\to\infty$.

Combining these $3$ estimates, we conclude that as $n\to\infty$
$$
\|q_n\|^2\ci{L^2(\mu)}=\int_{\T} |q_n|^2\,dm+\int_\D
|q_n|^2\,d\nu+\sum_k\mu_k|q_n(z_k)|^2\to 1+0+0=1\,.
$$
Now we would like to do something similar in the general case. The main
difficulty is that the Taylor series of the
function $F$ in general does not converge to $F$ uniformly on $\T$.
Fortunately, we do not really need the uniform
convergence here. Let us find out what kind of convergence it would be
appropriate to ask for.

Firstly, in order to have $\int_{\T}|q_n|^2w\,dm\to \int_{\T}|F|^2w\,dm$,
it
suffices to ensure that $S_n\to F$ in $L^2(w\,dm)$.

Secondly, let us estimate the (now possibly infinite) sum $\sum_k
\mu_k|q_n(z_k)|^2$. Assuming for a moment that $F\in H^2$, we can
try to estimate $|q_n(z_k)|^2$ by $\int_{\T} P_{z_k}|S_n-F|^2\,dm$ where
$P_{z_k}$ is the Poisson kernel corresponding to the point
$z_k$ (instead of the maximum principle, we use the subharmonicity of
$|z^n(S_n(z)-F(z))|^2$ in the exterior of the unit disk).
Therefore, to conclude that this sum goes to $0$, it suffices to ensure
that $S_n\to F$ in $L^2(w_1\,dm)$ where
$w_1=\sum_k\mu_k P_{z_k}\in L^1(m)$.

At last, to estimate $\int_\D |q_n|^2\,d\nu$, let us observe that the
assumption $F\in H^2$ is sufficient to ensure that $q_n(z)\to 0$
uniformly on compact subsets of $\D$ as $n\to\infty$ (the proof is exactly
the same as before). The dominated convergence
theorem is somewhat difficult to employ now because it would require an
$L^2$ estimate for the $\sup_n|S_n|$ on $\T$, i.e.,
a Carleson type theorem, but, fortunately, other boundedness conditions
are available to ensure that uniform convergence to $0$
on compact subsets of $\D$ implies convergence to $0$ in $L^2(\nu)$. The
simplest such condition is uniform boundedness of the
integrals $\int_\D|q_n|^2\,d\widetilde{\nu}$ where $\widetilde{\nu}$ is
any finite measure
of the kind $d\widetilde{\nu}=\varphi(|z|)\,d\nu$ with some positive
function $\varphi(r)$ increasing to $+\infty$ as $r\to 1-$. Indeed, then,
for
any $r\in(0,1)$, we can write
$$
\int_\D|q_n|^2\,d\nu=\int_{r\D}|q_n|^2\,d\nu+\int_{\D\setminus
r\D}|q_n|^2\,d\nu
\le
\max_{rD}|q_n|^2\nu(\D)+\frac{1}{\varphi(r)}\int_\D|q_n|^2\,d\widetilde{\nu}
$$
and observe that the first term tends to $0$ for any fixed $r\in(0,1)$ as
$n\to\infty$ while the second one can be made
arbitrarily small by choosing $r$ sufficiently close to $1$. Using the
subharmonicity of $|q_n|^2$ in $\D$, we see that to get
$\int_\D|q_n|^2\,d\nu\to 0$, it would suffice to have a uniform bound for
$\int_{\T} |S_n|^2w_2\,dm$ where
$w_2(z)=\int_\D P_\zeta(z)\,d\widetilde{\nu}(\zeta)$ is the ``harmonic
sweeping" of
the measure $\widetilde{\nu}$ to the unit circumference
$\T$. Note that, again, we have $w_2\in L^1(m)$.

The moral of the story is that it would suffice to ensure convergence of
$S_n$ to $F$ in $L^2(W\,dm)$ where $W=1+w+w_1+w_2$ is a
certain $L^1$ function on $\T$. (we added $1$ just to ensure that
$L^2(W\,dm)\subset L^2(m)$). Of course, we cannot hope for that
kind of convergence if the function $F$ itself is not in the space and, if
we define it exactly as before by $F=\psi B$, most likely,
it'll fail to be there. So let us see whether we can modify the definition
of $F$. Apparently, we cannot do anything with the
second factor: we need $F$ to vanish at all points $z_k$ in order to carry
out our trick in the estimate of $\sum_k\mu_k|q_n(z_k)|^2$.
On the other hand, after some thought, one can realize that we do not need
the first factor to be exactly $\psi$: any outer
function $\widetilde{\psi}$ with $|\widetilde{\psi}|\le|\psi|$ and
$\widetilde{\psi}(\infty)\approx\psi(\infty)$ will do just as well. This
freedom
allows us
to make $F$ belong to any given weighted space $L^2(V\,dm)$ with $V\ge 1$
(again, this condition is imposed just to get $F\in H^2$ for
sure) satisfying $\int_{\T}\log V\,dm<+\infty$. Indeed, just define
$\widetilde{\psi}$ by $|\widetilde{\psi}|^2=\min\left\{\frac 1w, \frac
AV\right\}$
on $\T$, $\widetilde{\psi}(\infty)>0$. By choosing $A$ sufficiently large,
we can
ensure that $\widetilde{\psi}(\infty)$ is as close to $\psi(\infty)$ as
we wish. On the other hand, we shall always have $\|F\|\ci{L^2(V\,dm)}=
\|\widetilde{\psi}\|\ci{L^2(V\,dm)}\le \sqrt A<+\infty$.

Since $W\in L^1(m)$ implies $\int_{\T}\log W\,dm<+\infty$, we, indeed, can
make $F\in L^2(W\,dm)$ by choosing $V$ equal to $W$ or any
larger weight with integrable logarithm. Unfortunately, this is not
enough. We need more, namely, that $S_n\to F$ in $L^2(W\,dm)$.
In particular, it implies that we must have a uniform bound for the norms
$\|S_n\|\ci{L^2(W\,dm)}$. Since $S_n=z^{-n}\cP_+(z^n F)$, where
$\cP_+$ is the orthogonal projection from $L^2(m)$ to $H^2$, we are
naturally led to the question for which integrable weights $W\ge 1$ one
can find another weight $V\ge W$ with $\log V\in L^1(m)$ such that $\cP_+$
is bounded as an operator from $L^2(V\,dm)$ to $L^2(W\,dm)$.
The answer is given by the celebrated Theorem of Koosis:

\begin{Koosisthm}
\label{Koosis}
For every integrable weight $W\ge 1$ one can find another weight $V\ge W$
with $\log V\in L^1(m)$ such that $\cP_+$ is bounded as an
operator from $L^2(V\,dm)$ to $L^2(W\,dm)$.
\end{Koosisthm}

\medskip

This is a truly remarkable theorem that deserves to be known much better
than it currently seems to be. For reader's convenience, we
included its proof in Appendix. Now let us finish the proof of our
theorem. The only remaining difficulty is that we need convergence
of $S_n$ to $F$ in $L^2(W\,dm)$ rather than mere boundedness of
$\|S_n\|\ci{L^2(W\,dm)}$, which is all the Koosis theorem provides us
with if we apply it directly to the weight $W$. The (fairly standard)
trick is to apply the Koosis theorem to another weight
$\widetilde{W}=W\varphi(W)$
where the increasing function $\varphi:[1,+\infty)\to[1,+\infty)$ is
chosen
so that $\lim_{x\to+\infty}\varphi(x)=+\infty$ and the weight
$\widetilde{W}$ is still integrable. Let now $V$ be the weight
corresponding to
$\widetilde{W}$ instead of just $W$. We claim that
$\|S_n-F\|\ci{L^2(W\,dm)}\to 0$ as $n\to\infty$ for all $F\in L^2(V\,dm)$.
Indeed, since $V\ge \widetilde{W}\ge 1$, we know that $F\in L^2(m)$
and, thereby, $\|S_n-F\|\ci{L^2(m)}\to 0$. On the other hand, the norms
$\|S_n-F\|\ci{L^2(\widetilde{W}\,dm)}$ are uniformly bounded.
Hence, for every $M>0$, we can write
$$
\int_{\T}|S_n-F|^2W\,dm=\int_{\{W\le M\}}|S_n-F|^2W\,dm + \int_{\{W>
M\}}|S_n-F|^2W\,dm
$$
$$
\le
M\int_{\T}|S_n-F|^2\,dm+\frac1{\varphi(M)}\int_{\T}|S_n-F|^2\widetilde{W}\,dm\,.
$$
Now, the first term tends to $0$ as $n\to\infty$ for any fixed $M>0$ while
the second one can be made arbitrarily small
by choosing $M$ large enough.

The theorem is thus completely proved.

\section{Proof of Corollary}
\label{proofasymp}

Again, denote by $B_\ell$ the partial Blaschke product with this zeros
$z_k$, $k\le\ell$.
Consider the integral
$$
\int_{\T} \left| 1 - \frac{p_n(z)}{z^n}
\frac{1}{\psi(z)B_\ell(z)}\right|^2\, dm\,.
$$
Using the residue theorem, we conclude that it equals
$$
1+\|p_n\|\ci{L^2(w\,dm)} - 2 \frac{ \tau_n }{\psi(\infty) B_\ell(\infty)}
- 2 \Re \sum_{k\,:\,k\le\ell}
\frac{p_n(z_k)}{z_k^{n+1} \psi(z_k) B_\ell'(z_k)}\,.
$$
We have already seen that $\ds \sum_{k\,:\,k\le\ell}
\frac{p_n(z_k)}{z_k^{n+1} \psi(z_k) B_\ell'(z_k)}\to 0$ as $n\to\infty$.
Also, $\|p_n\|\ci{L^2(w\,dm)}\le \|p_n\|\ci{L^2(\mu)}=1$.
Thus,
$$
\limsup_{n\to\infty}\int_{\T} \left|B_\ell(z) -
\frac{p_n(z)}{z^n\psi(z)}\right|^2 \,{dm}
\le 2\left(1-\frac{B(\infty)}{B_\ell(\infty)}\right)\,,
$$
whence
\begin{multline*}
\limsup_{n\to\infty}\int_{\T} \left| B(z) -
\frac{p_n(z)}{z^n\psi(z)}\right|^2 \,{dm}
\\
\le
2\|B_\ell-B\|\ci{L^2(m)}+4\left(1-\frac{B(\infty)}{B_\ell(\infty)}\right)\,.
\end{multline*}
Since the right hand side of the last inequality tends to $0$ as
$n\to\infty$, we conclude that
$$
\lim_{n\to\infty}\int_{\T} \left|B(z) - \frac{p_n(z)}{z^n\psi(z)}\right|^2
\,{dm}=0\,.
$$
It remains to recall that, for analytic in the exterior of the unit disk
functions
$\ds g_n(z)=B(z) - \frac{p_n(z)}{z^n\psi(z)}$, convergence to $0$ in $H^2$
is stronger than
pointwise convergence to $0$ in the exterior of the unit disk.

\section{Appendix: Proof of the Koosis theorem.}
\label{proofKoosis}

We shall outline the original proof from \cite{K} here. First of all,
note that for any two weights $V\ge W\ge 1$, the boundedness
of $\cP_+$ as an operator from $L^2(V)$ to $L^2(W)$ is implied by
(actually, equivalent to) its boundedness as an operator from
$L^2(w)$ to $L^2(v)$ where $w=\dfrac 1W$, $v=\dfrac 1V$. The latter is
understood in the sense that there exists a finite constant $C>0$
such that $\int_{\T}|\cP_+g|^2v\,dm\le C\int_{\T}|g|^2w\,dm$ for any
function
$g\in L^2(m)\cap L^2(w\,dm)$. This can be seen by a standard duality
argument. Using the density of trigonometric polynomials in $L^2(m)$, we
also see that it is enough to check this estimate
for the case when $g$ is a real trigonometric polynomial. Secondly, let us
note that $\cP_+g=\dfrac12(\widehat g(0)+g+i\widetilde{g})$ where
$\widetilde{\cdot}$ is the operator of harmonic conjugation, i.e., the
operator
that maps $\sum_j c_k z^k$ to
$\frac1i\sum_j (\operatorname{sgn} j) c_k z^k$. Since the identity
operator is bounded from $L^2(w\,dm)$ to $L^2(v\,dm)$ for any $v\le w$.
Since $|\widehat g(0)|=\left|\int_{\T} g\,dm\right|\le \left(\int_{\T}
|g|^2w\,dm\right)^{\frac12} \left(\int_{\T}
W\,dm\right)^{\frac12}=\sqrt{\|W\|\ci{L^1(m)}}\|g\|\ci{L^2(w\,dm)}$, we
see that the operator that maps $g$ to the constant function $\widehat
g(0)$ is bounded
in $L^2(w\,dm)$ and, thereby, from $L^2(w\,dm)$ to $L^2(v\,dm)$ for any
$v\le w$. These two remarks show that it is enough to construct
a weight $v\le w$ with integrable logarithm such that $\int_{\T}
|\widetilde{g}|^2v\,dm\le C \int_{\T} |g|^2w\,dm$ for any real
trigonometric
polynomial $g$ with $\widehat g(0)=0$. To this end, consider an outer
function $\Omega(z)$ with $\Re\Omega=W$ on $\T$. Note that
$$
\left|1-\frac
W{\Omega}\right|=\left|\frac{\Omega-\Re\Omega}{\Omega}\right|=\left|\frac{\Im\Omega}{\Omega}\right|<1\text{
almost everywhere on }\T\,.
$$
Let $\rho=1-\left|1-\frac W{\Omega}\right|$. Consider the analytic
polynomial $P(z)=g(z)+i\widetilde{g}(z)$. Since $P(0)=0$, we have
$$
\int_{\T} \frac{P^2}\Omega\,dm=\frac{P(0)^2}{\Omega(0)}=0\,.
$$
Let us rewrite it as
$$
\int_{\T} {P^2}w\,dm =\int_{\T} {P^2}\left(1-\frac W\Omega\right)w\,dm
$$
and take the real part of the left hand side with minus sign and the
absolute value of the right hand side. We shall get the inequality
$$
\int_{\T}(\widetilde{g}^2-g^2)w\,dm\le \int_{\T} |P|^2\left|1-\frac
W\Omega\right|w\,dm=
\int_{\T} (g^2+\widetilde{g}^2)(1-\rho)w\,dm\,,
$$
which is equivalent to
$$
\int_{\T} {\widetilde{g}^2}\rho w\,dm\le \int_{\T}  {g^2}(2-\rho) w\,dm\le
2\int_{\T}
{g^2}w\,dm\,.
$$
Thus, we can choose $v=\rho w$. The only thing that remains to check is
that $\int_{\T}\log v>-\infty$. To this end, note that
$$
\rho=1-\left|\frac{\Im \Omega}{\Omega}\right|\ge
\frac12\left(1-\left|\frac{\Im \Omega}{\Omega}\right|^2\right)
=\frac{|\Re \Omega|^2}{2|\Omega|^2}=\frac{W^2}{|\Omega|^2}\,.
$$
So $v=\rho w\ge \dfrac{W}{2|\Omega|^2}$. It remains to note that $W\ge 1$
while $\log|\Omega|\in L^1(m)$.

The Koosis theorem is thus completely proved.

\end{document}